\documentclass[aps,prl,twocolumn,superscriptaddress,showpacs]{revtex4} 

\usepackage{graphicx}
\usepackage{rotating}

\begin{document}


\newcommand{\fix}[1]{} 
\newcommand{\MET}{\mbox{$\raisebox{.3ex}{$\not\!$}E_T$}}
\newcommand{\METVEC}{\mbox{$\raisebox{.3ex}{$\not\!$}{\vec E}_T$}}
\def\DZero{D0~}



\title{Measurement of the Single Top Quark Production Cross Section at CDF}


\affiliation{Institute of Physics, Academia Sinica, Taipei, Taiwan 11529, Republic of China} 
\affiliation{Argonne National Laboratory, Argonne, Illinois 60439} 
\affiliation{University of Athens, 157 71 Athens, Greece} 
\affiliation{Institut de Fisica d'Altes Energies, Universitat Autonoma de Barcelona, E-08193, Bellaterra (Barcelona), Spain} 
\affiliation{Baylor University, Waco, Texas  76798} 
\affiliation{Istituto Nazionale di Fisica Nucleare Bologna, $^t$University of Bologna, I-40127 Bologna, Italy} 
\affiliation{Brandeis University, Waltham, Massachusetts 02254} 
\affiliation{University of California, Davis, Davis, California  95616} 
\affiliation{University of California, Los Angeles, Los Angeles, California  90024} 
\affiliation{University of California, San Diego, La Jolla, California  92093} 
\affiliation{University of California, Santa Barbara, Santa Barbara, California 93106} 
\affiliation{Instituto de Fisica de Cantabria, CSIC-University of Cantabria, 39005 Santander, Spain} 
\affiliation{Carnegie Mellon University, Pittsburgh, PA  15213} 
\affiliation{Enrico Fermi Institute, University of Chicago, Chicago, Illinois 60637} 
\affiliation{Comenius University, 842 48 Bratislava, Slovakia; Institute of Experimental Physics, 040 01 Kosice, Slovakia} 
\affiliation{Joint Institute for Nuclear Research, RU-141980 Dubna, Russia} 
\affiliation{Duke University, Durham, North Carolina  27708} 
\affiliation{Fermi National Accelerator Laboratory, Batavia, Illinois 60510} 
\affiliation{University of Florida, Gainesville, Florida  32611} 
\affiliation{Laboratori Nazionali di Frascati, Istituto Nazionale di Fisica Nucleare, I-00044 Frascati, Italy} 
\affiliation{University of Geneva, CH-1211 Geneva 4, Switzerland} 
\affiliation{Glasgow University, Glasgow G12 8QQ, United Kingdom} 
\affiliation{Harvard University, Cambridge, Massachusetts 02138} 
\affiliation{Division of High Energy Physics, Department of Physics, University of Helsinki and Helsinki Institute of Physics, FIN-00014, Helsinki, Finland} 
\affiliation{University of Illinois, Urbana, Illinois 61801} 
\affiliation{The Johns Hopkins University, Baltimore, Maryland 21218} 
\affiliation{Institut f\"{u}r Experimentelle Kernphysik, Universit\"{a}t Karlsruhe, 76128 Karlsruhe, Germany} 
\affiliation{Center for High Energy Physics: Kyungpook National University, Daegu 702-701, Korea; Seoul National University, Seoul 151-742, Korea; Sungkyunkwan University, Suwon 440-746, Korea; Korea Institute of Science and Technology Information, Daejeon, 305-806, Korea; Chonnam National University, Gwangju, 500-757, Korea} 
\affiliation{Ernest Orlando Lawrence Berkeley National Laboratory, Berkeley, California 94720} 
\affiliation{University of Liverpool, Liverpool L69 7ZE, United Kingdom} 
\affiliation{University College London, London WC1E 6BT, United Kingdom} 
\affiliation{Centro de Investigaciones Energeticas Medioambientales y Tecnologicas, E-28040 Madrid, Spain} 
\affiliation{Massachusetts Institute of Technology, Cambridge, Massachusetts  02139} 
\affiliation{Institute of Particle Physics: McGill University, Montr\'{e}al, Canada H3A~2T8; and University of Toronto, Toronto, Canada M5S~1A7} 
\affiliation{University of Michigan, Ann Arbor, Michigan 48109} 
\affiliation{Michigan State University, East Lansing, Michigan  48824}
\affiliation{Institution for Theoretical and Experimental Physics, ITEP, Moscow 117259, Russia} 
\affiliation{University of New Mexico, Albuquerque, New Mexico 87131} 
\affiliation{Northwestern University, Evanston, Illinois  60208} 
\affiliation{The Ohio State University, Columbus, Ohio  43210} 
\affiliation{Okayama University, Okayama 700-8530, Japan} 
\affiliation{Osaka City University, Osaka 588, Japan} 
\affiliation{University of Oxford, Oxford OX1 3RH, United Kingdom} 
\affiliation{Istituto Nazionale di Fisica Nucleare, Sezione di Padova-Trento, $^u$University of Padova, I-35131 Padova, Italy} 
\affiliation{LPNHE, Universite Pierre et Marie Curie/IN2P3-CNRS, UMR7585, Paris, F-75252 France} 
\affiliation{University of Pennsylvania, Philadelphia, Pennsylvania 19104} 
\affiliation{Istituto Nazionale di Fisica Nucleare Pisa, $^q$University of Pisa, $^r$University of Siena and $^s$Scuola Normale Superiore, I-56127 Pisa, Italy} 
\affiliation{University of Pittsburgh, Pittsburgh, Pennsylvania 15260} 
\affiliation{Purdue University, West Lafayette, Indiana 47907} 
\affiliation{University of Rochester, Rochester, New York 14627} 
\affiliation{The Rockefeller University, New York, New York 10021} 

\affiliation{Istituto Nazionale di Fisica Nucleare, Sezione di Roma 1, $^v$Sapienza Universit\`{a} di Roma, I-00185 Roma, Italy} 

\affiliation{Rutgers University, Piscataway, New Jersey 08855} 
\affiliation{Texas A\&M University, College Station, Texas 77843} 
\affiliation{Istituto Nazionale di Fisica Nucleare Trieste/\ Udine, $^w$University of Trieste/\ Udine, Italy} 
\affiliation{University of Tsukuba, Tsukuba, Ibaraki 305, Japan} 
\affiliation{Tufts University, Medford, Massachusetts 02155} 
\affiliation{Waseda University, Tokyo 169, Japan} 
\affiliation{Wayne State University, Detroit, Michigan  48201} 
\affiliation{University of Wisconsin, Madison, Wisconsin 53706} 
\affiliation{Yale University, New Haven, Connecticut 06520} 
\author{T.~Aaltonen}
\affiliation{Division of High Energy Physics, Department of Physics, University of Helsinki and Helsinki Institute of Physics, FIN-00014, Helsinki, Finland}
\author{J.~Adelman}
\affiliation{Enrico Fermi Institute, University of Chicago, Chicago, Illinois 60637}
\author{T.~Akimoto}
\affiliation{University of Tsukuba, Tsukuba, Ibaraki 305, Japan}
\author{M.G.~Albrow}
\affiliation{Fermi National Accelerator Laboratory, Batavia, Illinois 60510}
\author{B.~\'{A}lvarez~Gonz\'{a}lez}
\affiliation{Instituto de Fisica de Cantabria, CSIC-University of Cantabria, 39005 Santander, Spain}
\author{S.~Amerio$^u$}
\affiliation{Istituto Nazionale di Fisica Nucleare, Sezione di Padova-Trento, $^u$University of Padova, I-35131 Padova, Italy} 

\author{D.~Amidei}
\affiliation{University of Michigan, Ann Arbor, Michigan 48109}
\author{A.~Anastassov}
\affiliation{Northwestern University, Evanston, Illinois  60208}
\author{A.~Annovi}
\affiliation{Laboratori Nazionali di Frascati, Istituto Nazionale di Fisica Nucleare, I-00044 Frascati, Italy}
\author{J.~Antos}
\affiliation{Comenius University, 842 48 Bratislava, Slovakia; Institute of Experimental Physics, 040 01 Kosice, Slovakia}
% ...

\collaboration{CDF Collaboration\footnote{With visitors from $^a$Universiteit Antwerpen, B-2610 Antwerp, Belgium, 
$^b$Chinese Academy of Sciences, Beijing 100864, China, 
$^c$University of Bristol, Bristol BS8 1TL, United Kingdom, 
$^d$University of California Irvine, Irvine, CA  92697, 
$^e$University of California Santa Cruz, Santa Cruz, CA  95064, 
$^f$Cornell University, Ithaca, NY  14853, 
$^g$University of Cyprus, Nicosia CY-1678, Cyprus, 
$^h$University College Dublin, Dublin 4, Ireland, 
$^i$University of Edinburgh, Edinburgh EH9 3JZ, United Kingdom, 
$^j$Universidad Iberoamericana, Mexico D.F., Mexico, 
$^k$University of Manchester, Manchester M13 9PL, England, 
$^l$Nagasaki Institute of Applied Science, Nagasaki, Japan, 
$^m$University de Oviedo, E-33007 Oviedo, Spain, 
$^n$Queen Mary, University of London, London, E1 4NS, England, 
$^o$Texas Tech University, Lubbock, TX  79409, 
$^p$IFIC(CSIC-Universitat de Valencia), 46071 Valencia, Spain,
$^x$Royal Society of Edinburgh/Scottish Executive Support Research Fellow,
$^y$Istituto Nazionale di Fisica Nucleare, Sezione di Cagliari, 09042 Monserrato (Cagliari), Italy 
}}
\noaffiliation

\date{\today}

\begin{abstract}

We report a measurement of the single top quark production cross section
in 2.2 ~fb$^{-1}$ of $p{\bar{p}}$
collision data collected by the Collider Detector at Fermilab at $\sqrt{s}=1.96$~TeV.
Candidate events are classified as 
signal-like by three parallel analyses which use likelihood, matrix element, and neural
network discriminants. These results are combined in order to improve the 
sensitivity. 
We observe a signal consistent with the standard model prediction,
but inconsistent with the background-only model by 3.7 standard deviations with
a median expected sensitivity of 4.9 standard deviations.
We measure a cross section of $2.2^{+0.7}_{-0.6}(\mathrm{stat+sys})$~pb, extract the CKM 
matrix element value \mbox{$|V_{tb}|=0.88^{+0.13}_{-0.12} (\mathrm{stat+sys})\pm 0.07(\mathrm{theory})$}, 
and set the limit $|V_{tb}|>0.66$ at the 95\% C.L.  

\end{abstract}

\pacs{14.65.Ha, 13.85.Qk, 12.15.Hh, 12.15.Ji}  



\maketitle




The top quark was discovered by the CDF and \DZero collaborations 
in 1995~\cite{cdfd0topdisc} in the strong interaction $p{\bar{p}}\rightarrow t{\bar t}+X$.
Since then, a comprehensive program of measurements has brought more precise
knowledge of the top quark's mass, pair-production cross section,
and a number of its decay properties~\cite{pdg}. The evidence strongly suggests
that the particle observed in 1995 is the $SU(2)$ partner of the bottom quark
and that it decays nearly 100\% of the time into $Wb$ with 
a very short lifetime. The weak couplings of the top quark are less well constrained,
except that $|V_{tb}|^2\gg |V_{td}|^2+|V_{ts}|^2$ \cite{topR}. Requiring that the
$3\times 3$ Cabibbo-Kobayashi-Maskawa (CKM) matrix is unitary implies that $|V_{tb}|\simeq 1$~\cite{pdg}.
With a matrix of higher rank, though, $|V_{tb}|$ could be small without 
measurably changing the $t\rightarrow Wb$ branching ratio. 
Production of single top quarks provides a direct measurement of $|V_{tb}|$ and
a test of the $b$-quark content of the proton.  

\begin{figure}[t]
\includegraphics[width=\columnwidth]{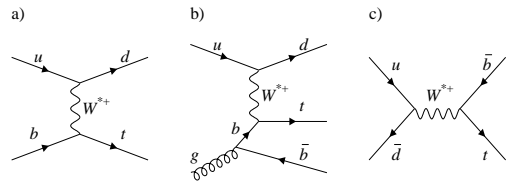}
\vskip -1.1cm
\caption{Representative Feynman diagrams of single top quark production. Figures (a) and (b)
are $t$-channel processes, and figure (c) is the $s$-channel process.}
\label{fig:sttfeyn}
\end{figure}

Top quarks are expected to be produced singly, as shown in Fig.~\ref{fig:sttfeyn}.  The
combined $s+t$-channel cross section is predicted at next-to-leading order (NLO)
to be $\sigma_{st}=2.86\pm 0.36$~pb~\cite{stxs}.
The small signal cross section and the presence of only one 
top quark in the final state make the separation of the signal from the large 
background challenging. Since the signal has very similar final
states to the standard model Higgs boson production process $WH\rightarrow \ell\nu b\bar{b}$, 
the methods of this analysis can be used to search for the Higgs boson.

Recently, the \DZero collaboration has reported evidence for
single top quark production using 0.9~fb$^{-1}$ of
data~\cite{d0evidence} while measuring a cross section of
$\sigma_{st}=4.7 \pm 1.3$~pb. 
This Letter reports a significantly more precise measurement of $\sigma_{st}$
in 2.2 fb$^{-1}$ of $p{\bar{p}}$ collisions at $\sqrt{s}=1.96$~TeV using the 
CDF II detector.

The CDF II detector~\cite{CDF} is a general purpose 
apparatus located at the Tevatron collider at Fermilab.
The detector consists of a solenoidal charged particle spectrometer which includes a silicon
microstrip detector array surrounded by a cylindrical drift chamber
in a 1.4~T axial magnetic field.  The energies of electrons and jets are measured with segmented
sampling calorimeters.
Surrounding the calorimeters are layers of steel instrumented with planar drift chambers and scintillators
used for muon identification.

Three distinct trigger algorithms are employed to select the data used in this analysis:
a high $p_T$ electron trigger, a high $p_T$ muon trigger, and a trigger 
that requires large missing transverse energy with either an energetic
electromagnetic cluster or two separated jets~\cite{coord,trigger}.


Events are further selected by requiring the presence of an isolated
electron or muon candidate with $p_T>20$~GeV$/c$,
large missing transverse energy $\MET>25$~GeV, and either two or three
jets each with $E_T>20$~GeV and $|\eta|<2.8$. The
jets are identified by a fixed-cone algorithm with radius $\Delta R
\equiv\sqrt{(\Delta \eta)^2+(\Delta \phi)^2} = 0.4$, and their
energies are corrected for instrumental
effects~\cite{jesnim}. At least one of the jets is required to have a
displaced vertex ($b$ tag) as identified by the {\sc secvtx} algorithm~\cite{secvtx}. 
This $b$ tag preferentially selects jets containing $B$ hadrons.

In order to reduce the $Z$+jets, $t\bar{t}$, and diboson backgrounds, 
candidate events with a second charged lepton are rejected. Cosmic ray and photon candidates are identified
and removed. Multijet background events without a leptonic $W$ decay 
(``non-$W$'') are rejected with specific selection requirements~\cite{Richter:2007zzc,pdongthesis}.


The diboson ({\it WW}, {\it WZ}, {\it ZZ}) and $t{\bar{t}}$ event yields are predicted using {\sc pythia}~\cite{pythia} 
Monte Carlo (MC) samples normalized to the theoretical cross section \cite{diboson,ttbarxs}.
The processes in which a vector boson is produced in association with one or more jets ($Z$ or $W$+jets) are generated
with {\sc alpgen} \cite{alpgen} using {\sc pythia}'s parton
showering and underlying event model.
The $W$+jets samples are normalized to
the measured data using events with exactly one, two, or three jets.  
A normalization factor of $1.4\pm0.4$ is applied to {\sc alpgen}'s prediction
for the fraction of $Wb\bar{b}$ and $W\!c\bar{c}$ events.
This factor is estimated by comparing the flavor content
in $b$-tagged $W$+1~jet events in the data to the prediction from simulation.
The background from events with mistakenly $b$-tagged light-flavor jets
({\it Wjj}\,) is estimated by measuring the rate of such mistags in multijet
data~\cite{secvtx}.
The mistag rate is then applied to the $W$+jets data samples before $b$ tagging.
The contributions to the data samples from non-$Wjj$ sources
are subtracted from the prediction~\cite{sarahsthesis}.
Multijet non-$W$ events typically have less $\MET$ than events containing $W$ bosons. 
By using templates for non-$W$ 
and $W$+jets, we fit the $\MET$ distribution and extract
the non-$W$ fraction in the high $\MET$ signal region.
The kinematic properties of the non-$W$ events are modeled using data events and $W$+jets
is modeled using MC simulated events.
The observed event yields and corresponding predictions 
are given in Table~\ref{tab:events}.  

\begin{table}[ht!]
\caption{Background composition and predicted number of single top events in 2.2~fb$^{-1}$ of
CDF Run II data with at least one $b$-tagged jet.}
\begin{center}
\begin{tabular}{lrclrcl}
\hline
\hline
Process & \multicolumn{3}{c}{$W+2$~jets} & \multicolumn{3}{c}{$W+3$~jets} \\
\hline
$s$-channel signal	& 40.3   & $\pm$ & 5.8   & 13.1  & $\pm$ & 1.9  \\
$t$-channel signal	& 60.8   & $\pm$ & 8.9   & 17.9  & $\pm$ & 2.6  \\
$Wb\bar{b}$		& 451.1  & $\pm$ & 136.0 & 138.0 & $\pm$ & 41.7 \\
$W\!c\bar{c}+W\!cj$	& 372.5  & $\pm$ & 114.8 & 103.2 & $\pm$ & 31.8 \\
{\it Wjj}		& 337.1  & $\pm$ & 41.9  & 101.6  & $\pm$ & 12.8 \\
$t\bar{t}$ 		& 142.0  & $\pm$ & 20.3  & 327.8  & $\pm$ & 46.6  \\
Non$-W$	  		& 60.5   & $\pm$ & 24.2  & 21.0  & $\pm$ & 8.4 \\
Diboson    		& 61.1   & $\pm$ & 6.2   & 20.4  & $\pm$ & 2.1  \\
$Z$+jets		& 25.5   & $\pm$ & 3.8   & 10.5   & $\pm$ & 1.5 \\
\hline					                       
Total prediction	& 1550.9 & $\pm$ & 256.6 & 753.5 & $\pm$ & 87.6  \\
\hline		
Observed  		& \multicolumn{3}{c}{1546}  &  \multicolumn{3}{c}{719}	         \\
\hline
\hline
\end{tabular}
\end{center}
\label{tab:events}
\end{table}


Single top events are simulated using the tree-level matrix-element
generator {\sc mad\-event}~\cite{madevent}. The two $t$-channel
processes of Figs.~\ref{fig:sttfeyn}(a) and \ref{fig:sttfeyn}(b) are
combined to match the event kinematics as predicted by a fully differential NLO
calculation~\cite{stxs,janlueck}.



The expected standard model signal-to-background ratio for selected events is
$\sim$7\% in the two-jet sample and $\sim$5\% in the three-jet sample.
The uncertainties on the background predictions are larger than
the expected signals; therefore, we have developed three powerful 
discriminants to distinguish signal from background events. The 
predicted distributions of each discriminant are fit to the 
data to extract the single top production cross
section. All analyses use the same event selection and were optimized
with the signal region blinded.

{\it Wjj}, $W\!c\bar{c}$ and $W\!cj$ events do not contain $b$-quark jets, but constitute $\sim$40\% of the
estimated background after imposing a $b$ tag requirement. As part of
all three discriminants we employ a jet-flavor separating variable, $b_{nn}$, constructed
using the neural network tool {\sc neurobayes}~\cite{Feindt:2006pm}, which is trained to distinguish
$b$ jets from charm and light-flavor jets based 
on secondary vertex tracking information~\cite{Richter:2007zzc}.
The usage of $b_{nn}$ leads to an 
improvement in sensitivity of 15 to 20\% in each analysis.

{\bf Likelihood Function Discriminant (LF):}
A projective likelihood technique ~\cite{sarahsthesis,lfcitation} is used to combine
information from several input variables to optimize the separation of the
single top signal from the backgrounds.  Two likelihood functions are created,
one for two-jet events, ${\cal{L}}_{2j}$, and one for three-jet events, ${\cal{L}}_{3j}$.  
The input variables used for ${\cal{L}}_{2j}$
are $b_{nn}$, $Q\times\eta$~\cite{cpyuan}, the invariant
mass of the $\ell\nu b$ system
$M_{\ell\nu b}$, the total scalar sum of transverse energy in the event $H_\mathrm{T}$, 
$\cos\theta^*_{\ell j}$~\cite{ctstar}, 
the dijet mass $M_{jj}$, and the $t$-channel matrix element.
The matrix element used here is computed using four-vectors from the event after
kinematically constraining $M_{\ell\nu}=M_W$ and $M_{\ell\nu b}=m_t$, where $M_W$ and
$m_t$ are the $W$ and top quark pole masses in the matrix element.  The $M_W$ constraint
introduces a quadratic ambiguity in the $z$ component of the neutrino momentum; we choose
the solution with the smaller $|p_z^\nu |$.

For events with three jets, ten input variables are used to construct ${\cal{L}}_{3j}$:
$b_{nn}$, 
$Q\times\eta$, 
$M_{\ell\nu b}$, 
$\cos\theta^*_{\ell j}$,
$M_{jjn}$ of the two jets not selected as the $b$ from top decay, 
the number of $b$ tags, 
the smallest $\Delta R$ between any two jets,
the smallest $p_T$ of the three jets, 
$p_T(W)$, 
and $E_T$ of the jet selected as coming from the $b$ from the top quark decay. The $b$-quark jet is
chosen using a linear combination of the jet $E_T$ and the $\chi^2$ resulting from the
comparison of the kinematically constrained jet energy and the measured jet energy.

{\bf Matrix Element Discriminant (ME):}
The matrix element method relies on the evaluation of event
probability densities for signal and background processes based on
calculations of the standard model differential
cross sections~\cite{pdongthesis,refME}. 
 We construct these probability densities for each process for each event given their measured quantities $x$ by
 integrating the appropriate differential cross section $d\sigma(y)/dy$ over the underlying partonic quantities $y$,
 convolved with the parton distribution functions (PDFs) and detector resolution effects:
\begin{equation}
\label{eq:prob}
P({x})  = \sum_{\mathrm{perm.}} \int  \frac{d\sigma(y)}{dy} f(q_1) f(q_2) dq_1 dq_2 W({x},{y})dy.
\end{equation}

\noindent 
The PDFs ($f(q_1)$ and $f(q_2)$) take into account the flavors 
of the colliding quark and anti-quark. We use the CTEQ PDF 
parameterization \cite{CTEQ}. The detector resolution effects are described by a
transfer function $W({x},{y})$ relating $x$ to $y$. 
The momenta of electrons, muons, and the angles of jets are assumed to be
measured exactly.
$W({x},{y})$ maps parton energies to measured jet energies after
correction for instrumental effects~\cite{jesnim}. This
mapping is obtained by parameterizing the jet response in fully
simulated MC events. 
 The definition 
of the probability densities includes possible permutations of matching 
jets with partons. 
The integration is performed
over the energy of the partons and $p_z^\nu$.
We calculate the matrix element for the event probability at
tree-level using 
{\sc madgraph}~\cite{madgraph}. Event probability densities 
are computed for the $s$-channel and $t$-channel signal as well as $Wb\bar b$,
$W\!c\bar c$, $W\!cj$, {\it Wjj}, and $t\bar{t}$ background hypotheses.
In the specific
case of the $t\bar{t}$ matrix element, additional integrations
are performed over the momenta of particles not detected.

The event probability densities are combined into an event
probability discriminant:
$EPD=P_{\mathrm{signal}}/(P_{\mathrm{signal}}+P_{\mathrm{background}})$.
To better classify signal events that contain $b$ jets,
we incorporate the output $b_{nn}$ of the neural network
jet-flavor separator into the final discriminant:
\begin{footnotesize} 
\begin{equation}
EPD=\frac{b_{nn} \cdot P_{st}}{b_{nn}(P_{st}\!+\!P_{t\bar t}\!+\!P_{Wb\bar b})\!+\!(1\!-\!b_{nn})(P_{Wc\bar c}\!+\!P_{Wcj}\!+\!P_{Wjj})}.
\nonumber
\end{equation}
\end{footnotesize}
Both signal channels are combined to one single top probability density
$P_{st}=P_{s-{\mathrm{channel}}}+P_{t-{\mathrm{channel}}}$. 

{\bf Neural Network Discriminant (NN):}
The third multivariate approach~\cite{Richter:2007zzc} employs neural networks, which have the general 
advantage that correlations between the discriminating input variables are 
identified and utilized to optimize the separation power between 
signal and background. The networks are developed using the 
{\sc neurobayes} analysis package~\cite{Feindt:2006pm}, which
combines a three-layer feed-forward neural network with a complex and robust 
preprocessing of the input variables. 
Bayesian regularization techniques are utilized to avoid over-training.

Four separate networks are trained to identify 
different signals in distinct samples using simulated events from
the common samples described previously. An $s$-channel signal is used for 
training on events with two $b$-tagged jets. A $t$-channel signal is 
used for the two-jet sample with a single $b$ tag and for the three-jet samples 
with one or two $b$ tags.
The networks use 11 to 18 input variables. The most 
important ones are $M_{\ell\nu b}$, $b_{nn}$, $M_{jj}$, 
$Q\times\eta$, 
$\cos\theta^*_{\ell j}$,
the transverse mass of the $W$ boson, 
and $H_\mathrm{T}$. The input variables are 
selected from a large list using an automated evaluation
during the preprocessing step before the network training.
In an iterative process, we determine those variables whose removal
would cause a significant loss in separation power between signal and 
background and use them for network training.

{\bf Combination:}
We studied two methods to combine the cross section fit results. 
The best linear unbiased estimator
(BLUE)~\cite{blue} technique optimizes the coefficients of a linear combination 
using the uncertainties and correlations of the three individual analyses: LF, ME, and NN.
The correlation coefficients between the analyses are: LF-ME:~59\%; LF-NN:~74\%; ME-NN:~61\%.
In another combination approach, a ``super analysis'' is built based on the 
outcomes for each event in each of the three individual analyses.
The super-discriminant method uses a neuro-evolution network~\cite{neat} trained
to separate the signal from the background based on the discriminant outputs of
the three analyses. 
With the super-discriminant analysis we improve the sensitivity
by 10\% over the best individual analysis, and we use it to quote our final results.  
As a cross-check, BLUE yields a 7\% sensitivity improvement.

Before unblinding the data, the MC simulation of each input variable and
the discriminant outputs were checked in data control samples depleted in signal.
These are the lepton + $b$-tagged four-jet sample, which is enriched in $t\bar{t}$ events, and
the two- and three-jet samples in which no jet is $b$ tagged.
The latter are high-statistics samples with similar kinematics to the $b$-tagged signal
samples. The data distributions in the control samples are described well by the MC simulation.

\begin{figure}[h]
\includegraphics[width=0.9\columnwidth]{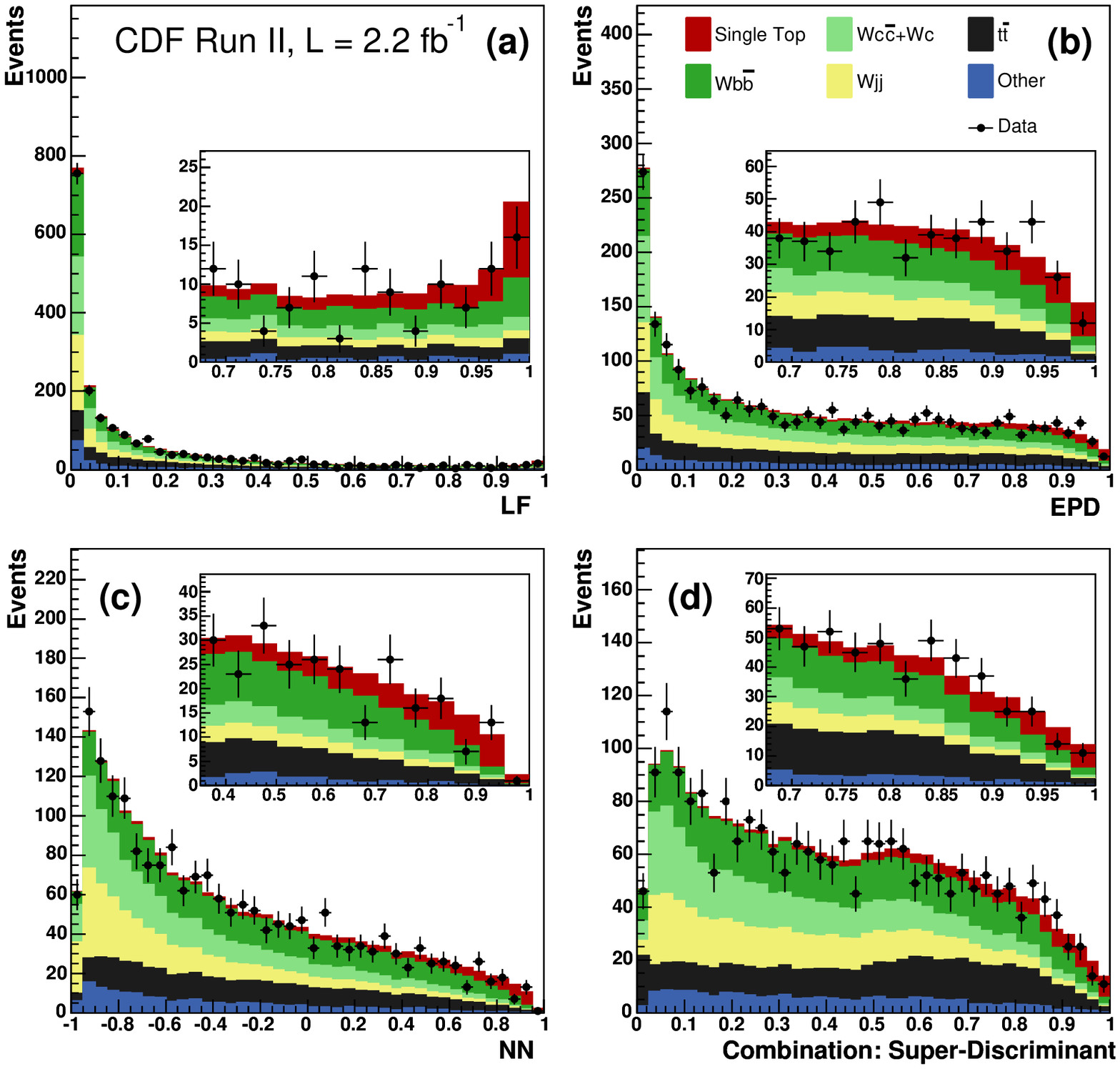}
\includegraphics[width=0.9\columnwidth]{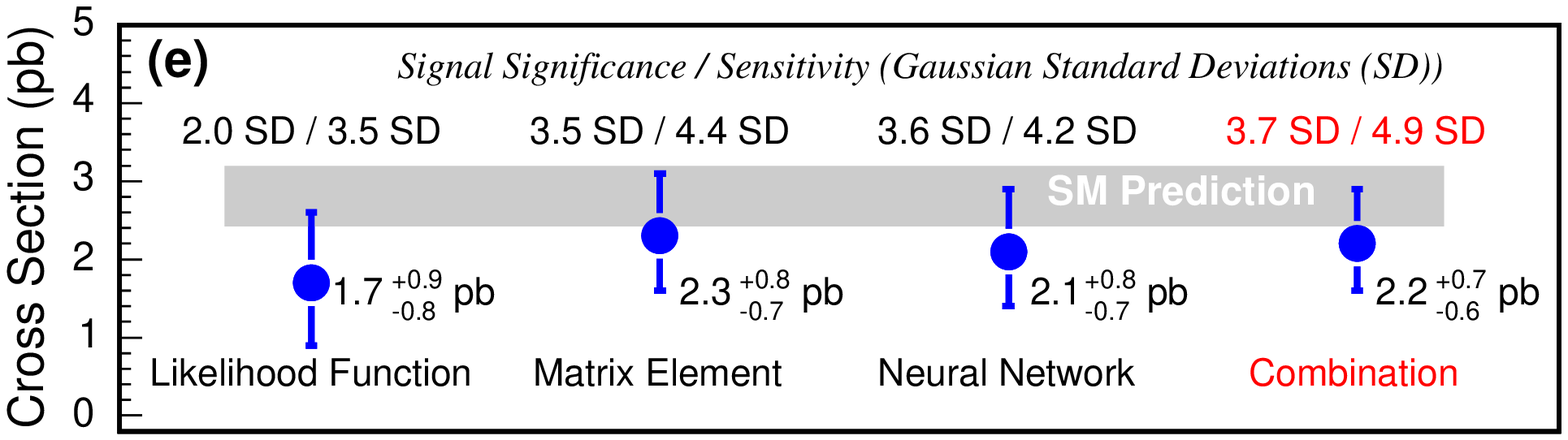}
\caption{Discriminant distribution for all channels combined of the
(a) LF analysis, (b) ME analysis, (c) NN
analysis and (d) combined super-discriminant analysis.  Points with error bars indicate
the data. The predicted signal and background distributions are shown
as stacked histograms. The insets show the candidate events in the signal regions.
A summary of all results is shown in (e).  
}
\label{fig:superdiscriminant}
\end{figure}

Figure~\ref{fig:superdiscriminant} shows the distributions of the 
individual analyses' discriminants and the super-discriminant.
We calculate the probability ($p$-value)~\cite{pdg} of the background-only discriminant distribution to 
fluctuate to the observed data or more which is then converted into signal 
significance under a Gaussian assumption.
All sources of systematic uncertainty are 
included and correlations between normalization
and discriminant shape changes are considered.
Uncertainties in the jet energy scale, $b$-tagging efficiencies, lepton identification
and trigger efficiencies, the amount of initial and final
state radiation, PDFs, factorization and renormalization scale,
and MC modeling have been explored and incorporated in this combination
and all individual analyses. 

We interpret the excess of signal-like events over the expected 
background as strong evidence for single top production with a signal significance of 3.7 
standard deviations, with a sensitivity, defined to be the median expected significance, of 4.9 standard deviations.
The most probable value of the
combined $s$-channel and $t$-channel cross sections is $\sigma_{st}=2.2^{+0.7}_{-0.6}$~pb
for a top quark mass of 175 GeV/$c^2$ which is consistent with the cross check result,
obtained from BLUE, $\sigma_{st}=2.3\pm 0.7$~pb.
The dependence on the top quark mass is +0.02~pb/(GeV/$c^2$).
From the cross section measurement at $m_t=$~175 GeV/$c^2$, we obtain
$|V_{tb}|=0.88^{+0.13}_{-0.12} (\mathrm{stat.+sys.})\pm 0.07$(theory\cite{stxs}) and limit $|V_{tb}|>0.66$
at the 95\% C.L. assuming a flat prior in $|V_{tb}|^2$ from 0 to 1.
This is the most precise direct measurement of $|V_{tb}|$ to date.

\begin{acknowledgments}

We thank the Fermilab staff and the technical staffs of the
participating institutions for their vital contributions. This work
was supported by the U.S. Department of Energy and National Science
Foundation; the Italian Istituto Nazionale di Fisica Nucleare; the
Ministry of Education, Culture, Sports, Science and Technology of
Japan; the Natural Sciences and Engineering Research Council of
Canada; the National Science Council of the Republic of China; the
Swiss National Science Foundation; the A.P. Sloan Foundation; the
Bundesministerium f\"ur Bildung und Forschung and the
Alexander von Humboldt Foundation, Germany; the Korean
Science and Engineering Foundation and the Korean Research Foundation;
the Science and Technology Facilities Council and the Royal Society,
UK; the Institut National de Physique Nucleaire et Physique des
Particules/CNRS; the Russian Foundation for Basic Research; the
Ministerio de Educaci\'{o}n y Ciencia and Programa Consolider-Ingenio
2010, Spain; the Slovak R\&D Agency; and the Academy of Finland.

\end{acknowledgments}


\begin{thebibliography}{99}
\bibitem{coord} We use a cylindrical coordinate system with its origin in the center 
of the detector, where $\theta$ and $\phi$ are the polar and azimuthal angles, respectively,
and pseudorapidity is $\eta= - \rm ln$ $\rm tan(\theta/2)$. 
The missing $E_T$ ($\METVEC$) is defined by $\METVEC = -\sum_iE_T^i{\hat n}_i, i={\rm calorimeter~tower~number}$,
where ${\hat n}_i$ is a unit vector perpendicular to the beam axis and pointing at the
$i^{th}$ calorimeter tower.  $\METVEC$ is corrected for high-energy muons and also jet energy corrections.
We define $\MET=|\METVEC|$. The transverse momentum $p_T$ is defined to be $p\sin\theta$.
\bibitem{cpyuan} $Q\times\eta$ is
the charge of the lepton times the pseudorapidity of the jet not assigned to be the b from top decay.
  C.~P.~Yuan,
  Phys.\ Rev.\  D {\bf 41}, 42 (1990).
\bibitem{ctstar}$\cos\theta^*_{\ell j}$ 
is the cosine of the angle between the lepton and the jet not assigned to be the b from top decay, 
in the top quark rest frame.
  G.~Mahlon and S.~J.~Parke,
  Phys.\ Rev.\  D {\bf 55}, 7249 (1997).
\end{thebibliography}
\end{document}